\documentclass[twocolumn,amsmath,amssymb,prl,superscriptaddress]{revtex4-1}

\usepackage{graphicx}
\usepackage{color}
\usepackage{dcolumn}
\usepackage{bm}
\usepackage{amsmath}
\usepackage{amsfonts}
\usepackage{amssymb}
\usepackage[nice]{nicefrac}
\usepackage[tight]{units}
\usepackage{MnSymbol}
\usepackage{amssymb,amsmath}

\def\t0{\mbox{$t_{\mbox{{\tiny {0}}}}$}}
\def\p0{\mbox{$p_{\mbox{{\tiny {0}}}}$}}
\def\E0{\mbox{$E_{\mbox{{\tiny {0}}}}$}}
\def \of#1{\!\left(#1\right)}

\begin{document}

\title{Probing single-photon ionization on the attosecond time scale}

\author{K.~Kl\"under}
\affiliation{Department of Physics, Lund University, P.O. Box 118, 22100 Lund, Sweden}

\author{J.~M.~Dahlstr\"om}
\affiliation{Department of Physics, Lund University, P.O. Box 118, 22100 Lund, Sweden}

\author{M.~Gisselbrecht}
\affiliation{Department of Physics, Lund University, P.O. Box 118, 22100 Lund, Sweden}

\author{T.~Fordell}
\affiliation{Department of Physics, Lund University, P.O. Box 118, 22100 Lund, Sweden}

\author{M.~Swoboda}
\affiliation{Department of Physics, Lund University, P.O. Box 118, 22100 Lund, Sweden}

\author{D.~Gu\'enot}
\affiliation{Department of Physics, Lund University, P.O. Box 118, 22100 Lund, Sweden}

\author{P.~Johnsson}
\affiliation{Department of Physics, Lund University, P.O. Box 118, 22100 Lund, Sweden}

\author{J.~Caillat}
\affiliation{Laboratoire de Chimie Physique-Mati\`ere et Rayonnement, Universit\'e Pierre et Marie Curie, 11, Rue Pierre et Marie Curie, 75231 Paris Cedex, 05, France}

\author{J.~Mauritsson}
\affiliation{Department of Physics, Lund University, P.O. Box 118, 22100 Lund, Sweden}

\author{A.~Maquet}
\affiliation{Laboratoire de Chimie Physique-Mati\`ere et Rayonnement, Universit\'e Pierre et Marie Curie, 11, Rue Pierre et Marie Curie, 75231 Paris Cedex, 05, France}

\author{R.~Ta\"ieb}
\affiliation{Laboratoire de Chimie Physique-Mati\`ere et Rayonnement, Universit\'e Pierre et Marie Curie, 11, Rue Pierre et Marie Curie, 75231 Paris Cedex, 05, France}

\author{A.~L'Huillier}
\email{anne.lhuillier@fysik.lth.se}
\homepage{http://www.atto.fysik.lth.se}
\affiliation{Department of Physics, Lund University, P.O. Box 118, 22100 Lund, Sweden}
\date{\today}
\begin{abstract}
We study photoionization of argon atoms excited by attosecond pulses using an interferometric measurement technique. We measure the difference in time delays between electrons emitted from the $3s^2$ and from the $3p^6$ shell, at different excitation energies ranging from 32 to 42 eV. The determination of single photoemission time delays requires to take into account the measurement process, involving the interaction with a probing infrared field. This contribution can be estimated using an universal formula and is found to account for a substantial fraction of the measured delay.
\end{abstract}
\maketitle
\thispagestyle{empty}
\noindent The interaction of light with matter is an essential process in nature and its paradigm, the photoelectric effect, has been studied during decades using synchrotron radiation~\cite{SchmidtRPP1992}. The development of ultrashort light pulses in the attosecond range allows scientists to tackle temporal aspects of electron transitions in atoms, molecules and more complex systems. Cavalieri {\it et al.}~\cite{CavalieriNature2007} investigated photoemission from the valence and the conduction band in tungsten crystals using single attosecond pulses and an infrared probing field through the streaking technique~\cite{SansoneScience2006,GoulielmakisScience2008}. Recently, Schultze {\it et al.}~\cite{SchultzeScience2010} implemented the same technique to study photoemission from the $2s^2$ and $2p^6$ shells in neon at a pulse energy of 100 eV. They measured a difference in photoemission time delays equal to 21~as, a value which is significantly larger than the expected theoretical value, as further discussed in a series of theoretical articles~\cite{YakovlevPRL2010,BaggesenPRL2010,ZhangPRA2010,KheifetsPRL2010}.\\
In this letter, we examine photoemission of electrons from the $3s^2$ and $3p^6$ shells in argon. Our method uses a frequency comb of high-order harmonics with photon energies varying from 32 to 42~eV for the photoionization and a weak infrared laser field for probing the outgoing electrons in time. Our method is complementary to the streaking method used in~\cite{SchultzeScience2010}. It is based upon interferometry and it allows us to explore the threshold region for the $3s^2$ shell, where one expects large variation in photoemission times. We measure a delay between the ionization from the $3s^2$ and $3p^6$ shells which varies with photon energy. We investigate the influence of the interaction with the weak infrared field, which is needed to get the interferometric measurement and therefore the temporal information. Probing the outgoing wave packet with even a weak infrared field affects the electron motion and therefore the measured delay. Fortunately, this effect can be analytically calculated and takes a universal form, which allows us to disentangle the different effects and gives us access to the single-photon ionization time, also called Wigner time~\cite{WignerPR1955,CarvalhoPhysRep2002}.

\begin{figure}[b]
\centering
\includegraphics[width=1\linewidth]{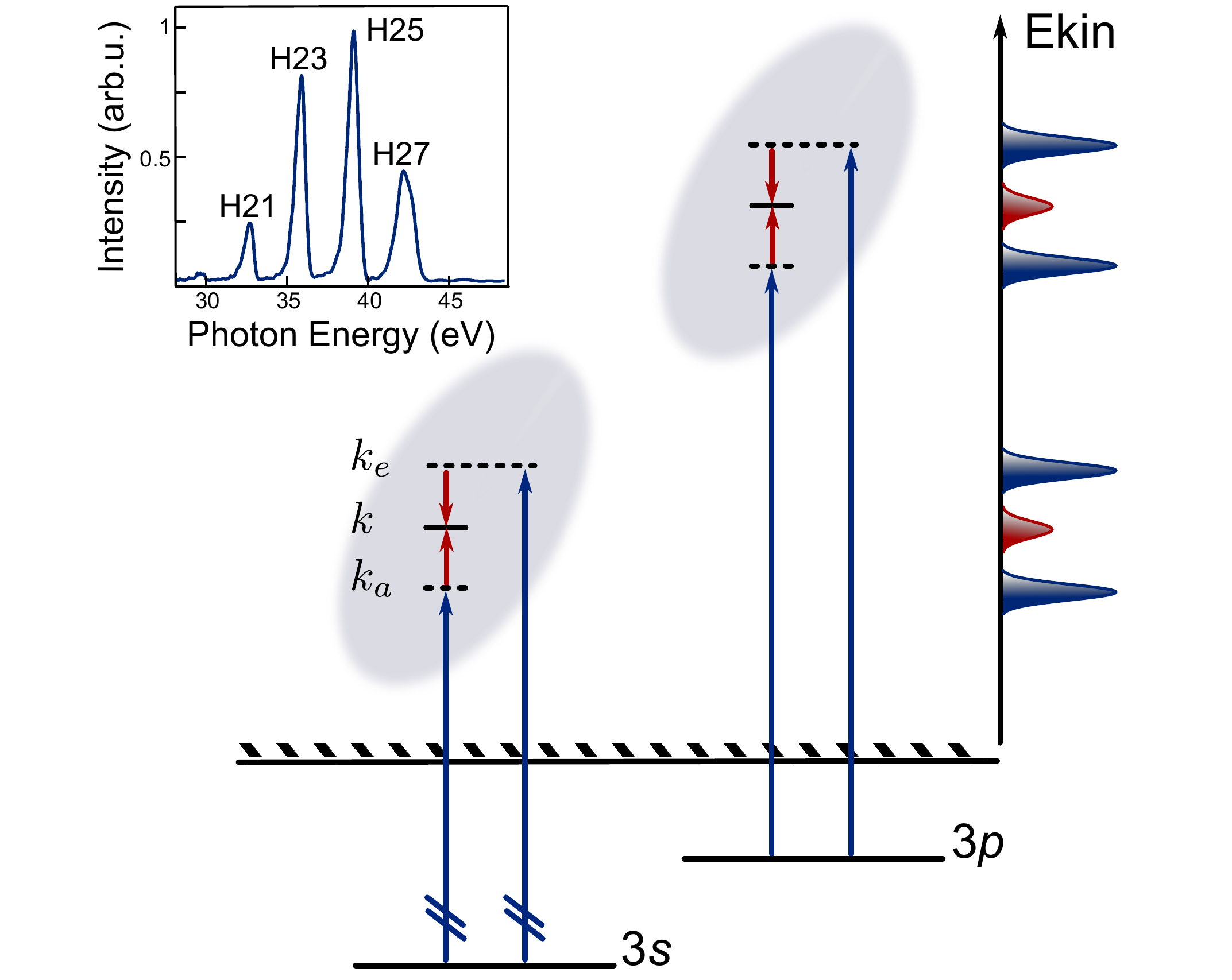}
\caption{(color online). Principle of the measurement. Two electron wave packets originating from different subshells are simultaneously created using the same comb of high order harmonics. The outgoing electron wave packets are further probed with a weak infrared field. For simplicity only two harmonics are indicated. Also shown is the experimental harmonic spectrum used.}
\label{fig:1}
\end{figure}
\noindent The basic principle of our experiment is shown in Fig.~\ref{fig:1}. We ionize argon using a comb of high-order harmonics. For a central frequency of the harmonic comb above the binding energy of the 3$s$ orbital we simultaneously create two independent electron wave packets, one originating from the 3$s^2$ and one from the 3$p^6$ subshell. The presence of a fraction of the fundamental laser field with frequency $\omega$ induces the formation of sideband peaks due to two-photon transitions including absorption or emission of an infrared photon~\cite{SchinsJOSA1996,VeniardPRA1996}.
Two different and interfering quantum paths involving consecutive harmonics lead to the same sideband (see Fig.~\ref{fig:1}). When changing the delay $\tau$ between the harmonic comb and the laser field, the sideband signal from a given subshell is modulated as~\cite{PaulScience2001}
\begin{equation}
S(\tau) = \alpha + \beta \cos [ 2\omega(\tau + \tau_\textsc{a}+\tau_\textsc{i})],
\label{eq:RABITT}
\end{equation}
where $\alpha,\beta$ are two constants independent of $\tau$. The term $\tau_\textsc{a}$ is proportional to the difference in phase between consecutive harmonics and describes the group delay of the attosecond pulses, while $\tau_\textsc{i}$ represents the atomic delay due to the two-photon ionization process~\cite{TomaJPB2002,HaesslerPRA2009,SwobodaPRL2010}. As we will show below $\tau_\textsc{i}$ can be connected to the Wigner time delay $\tau_\textsc{w}$ for the single-photon ionization. The knowledge of $\tau_\textsc{a}$ as well as of the absolute value of the delay $\tau$ would enable us to determine $\tau_\textsc{i}$ directly. However, these are difficult to obtain experimentally. The simultaneous measurement of the two electron wave packets allows us to cancel the influence of the attosecond group delay $\tau_\textsc{a}$ and to determine $\tau_{\textsc{i}}(3s)-\tau_{\textsc{i}}(3p)$ at the same photon energy, i.e., at kinetic energies separated by the difference in binding energy between the two orbitals (13.5 eV).\\
Our experiments were performed with a 800-nm 30-fs Titanium-Sapphire laser system~\cite{FordellOptExp2009}. High-order harmonics were generated in a pulsed Ar gas cell and spatially filtered using a small aperture~\cite{LopezPRL2005}. We used a 0.2~$\mu$m thick chromium thin film to select a 10 eV-broad spectral window corresponding to harmonic 21 to 27 at 38~eV central energy (see also Fig.~\ref{fig:1}). This filter was chosen to separate the wave packets emitted from the $3s$ and $3p$ subshells in energy. The comb of about four phase-locked harmonics, corresponding to a train of attosecond pulses with a duration of 450~as, was focused by a toroidal mirror into the sensitive region of a magnetic bottle electron spectrometer containing a diffusive Ar gas jet. Part of the laser field was extracted prior to the high-order harmonic generation and recombined downwards collinearly with the harmonics with a variable time delay $\tau$. The precision of our measurement does not depend on the duration of the attosecond XUV pulses but on the interferometric stability of our experiment.\\

\begin{figure}
\centering
\includegraphics[width=1\linewidth]{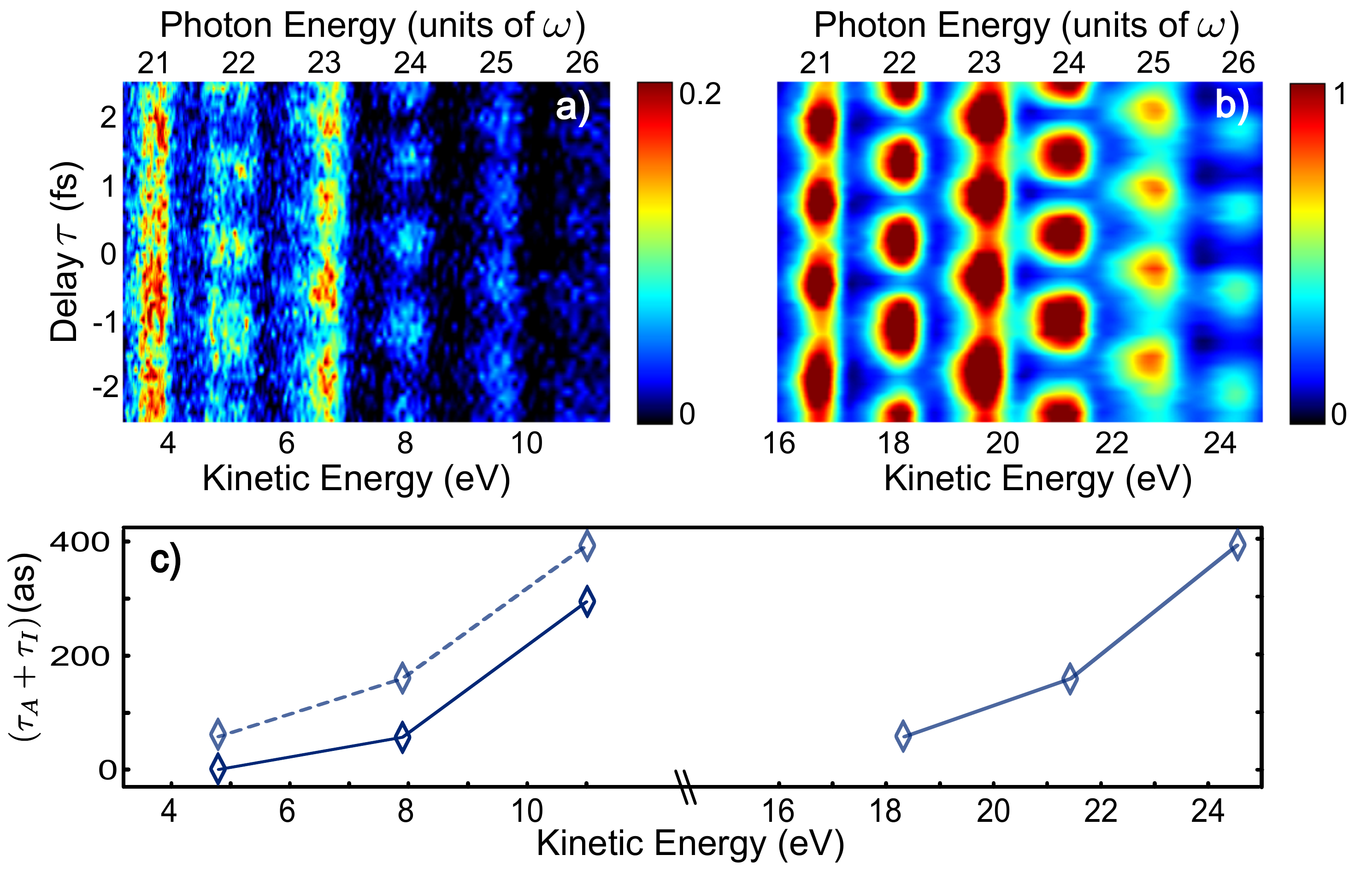}
\caption{(color online). Energy spectra as a function of delay from electrons liberated from the 3$s$ orbital (a) and the 3$p$ orbital (b), respectively. (c) Retrieved delays corrected for the Cr group delay. Also shown are the $3p$-delays shifted down in energy for comparison with the $3s$-delays (dashed line).}
\label{fig:2}
\end{figure}
\noindent Fig.~\ref{fig:2}(a) and 2(b) present electron spectra as a function of the delay $\tau$ between the XUV and the infrared pulses. The low-energy spectrum in Fig.~\ref{fig:2}(a) shows electron peaks at energies corresponding to single photon ionization from the $3s$ shell by the harmonics and additional sideband peaks due to two-photon transitions. The high-energy part of the spectrum shown in Fig.~\ref{fig:2}(b) presents the corresponding photoelectron spectra for $3p$ ionization.  Although simultaneously recorded the results are presented separately due to the unequal signal strength caused by the difference in cross section and detector sensitivity (note the different color scales). For both channels the sideband signal oscillates. Fig.~\ref{fig:2}(c) presents the delays obtained by Fourier transform of the sideband signal along the delay axis for the scan shown in (a) and (b), corrected for the influence of the Cr filter, which is positively dispersive in this region~\cite{XRayDatabase}.  The variation in delay reflects mainly the positive chirp of the attosecond pulses. The main experimental result of the present work is the significant offset between the delays measured for the two wave packets. To emphasize this result, we show as a dashed line the $3p$-delays shifted down in energy by 13.5~eV. Taking the difference between the measured delays at the same excitation energy and averaging over five independent measurements, we determine a difference in delays $\tau_{\textsc{i}}(3s)-\tau_{\textsc{i}}(3p)$ equal to -40$\pm$10~as for sideband~22, -110$\pm$10~as for sideband~24 and -80$\pm$30~as for sideband~26.

\begin{figure}
\centering
\includegraphics[width=1\linewidth]{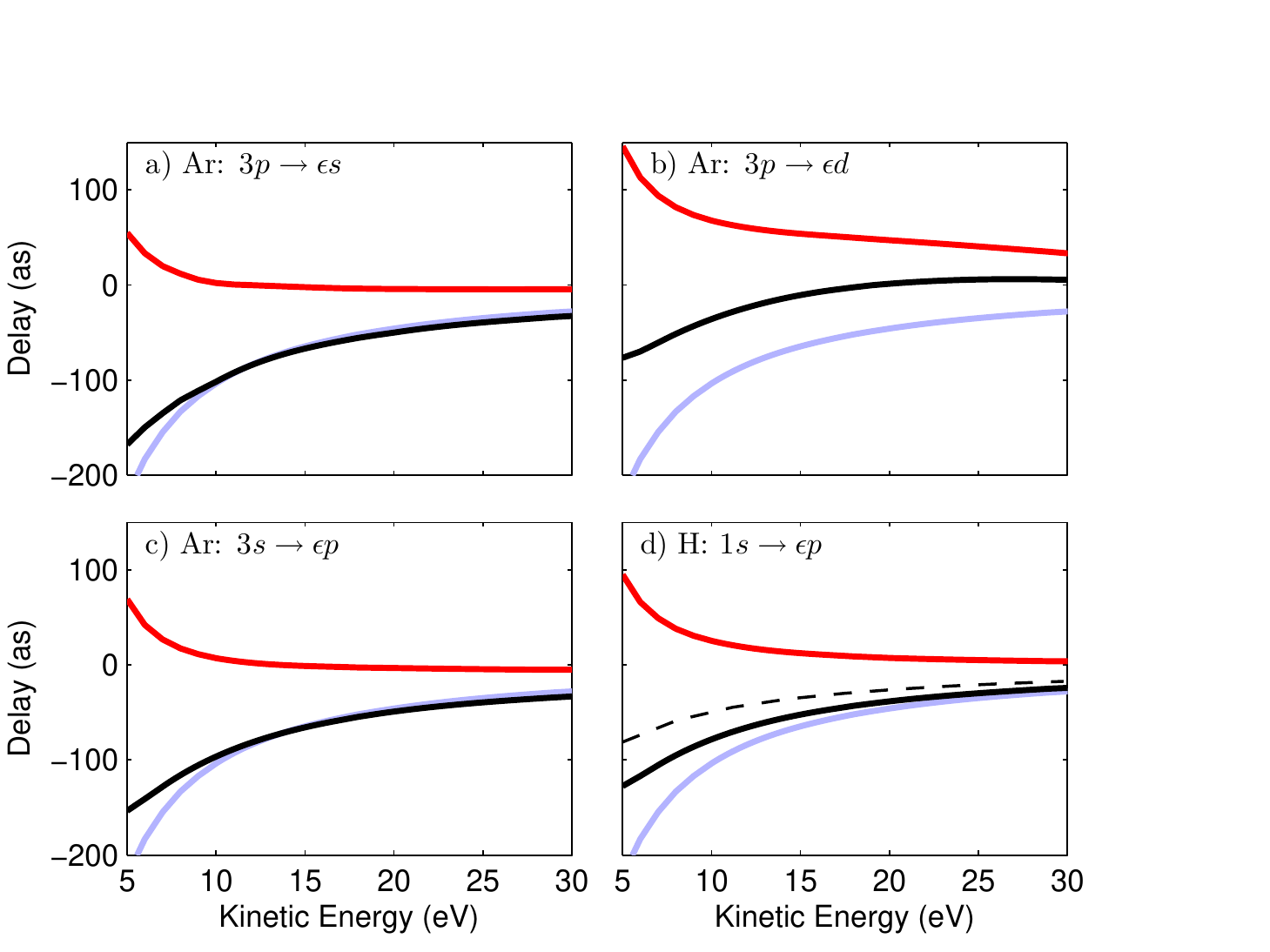}
\caption{(color online). Computed delays associated with the following ionization channels: (a) $3p\rightarrow \epsilon s$, (b) $3p\rightarrow \epsilon d$, (c) $3s\rightarrow \epsilon p$ in Ar, and (d) $1s\rightarrow \epsilon p$ in H. The red lines are the one-photon Wigner time delays. The blue lines represent the estimated delays induced by the measurement, $\tau_{\textsc{cc}}$. The sum of the two delays is shown as a black line. The dashed line in (d) is the result of an exact calculation in H.}
\label{fig:3}
\end{figure}

\begin{figure}
\centering
\includegraphics[width=1\linewidth]{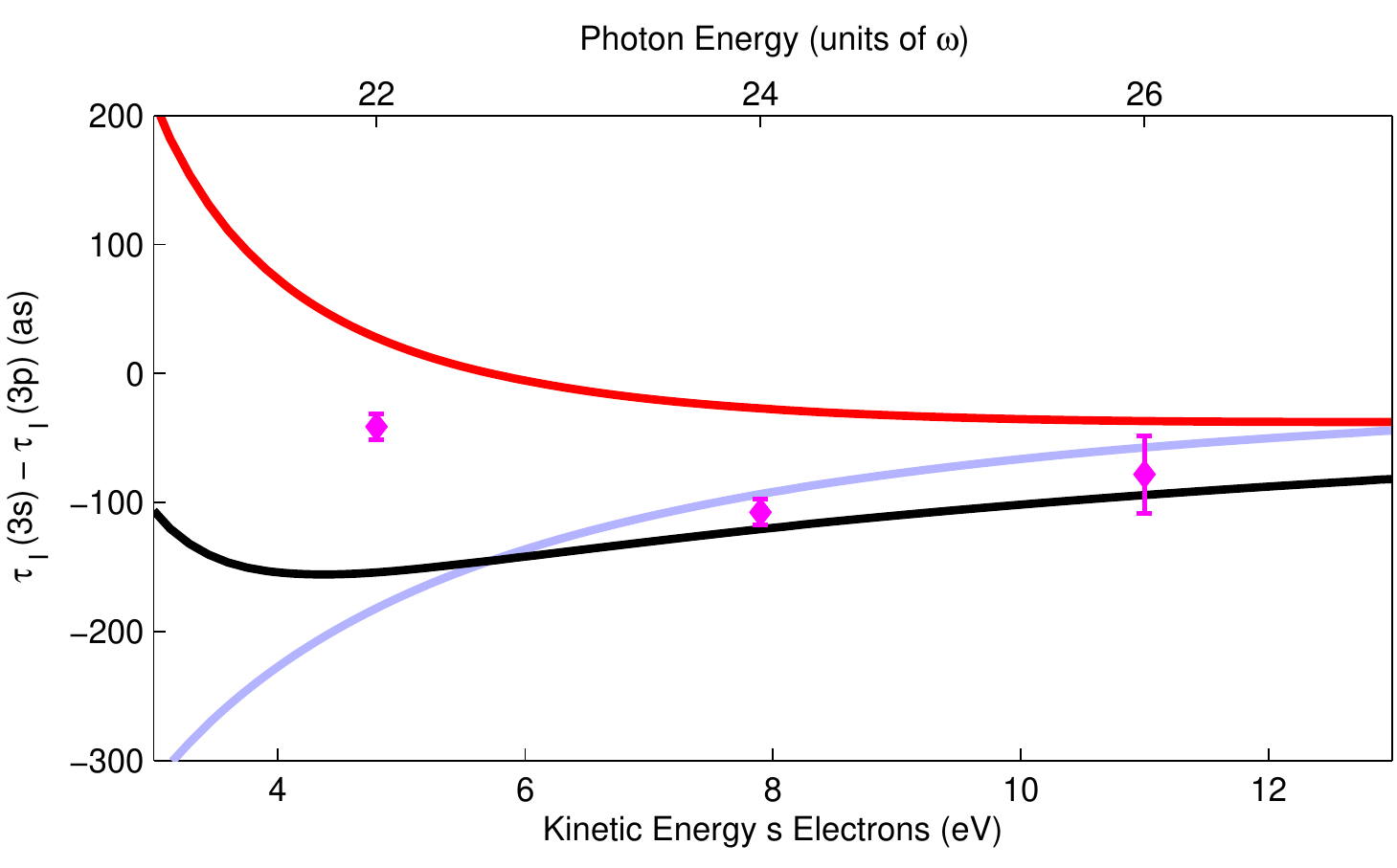}
\caption{Comparison between the measured delay differences for ionization of Ar from the $3s$ and $3p$ shells (diamonds) with calculations performed according to the approximate theory developed in this work (black line). Also shown is the delay expected for one-photon ionization (red line) and the laser driven continuum-continuum transition (blue line).}
\label{fig:4}
\end{figure}

\noindent To understand the meaning of these time delays, we need to establish the connection between single-photon ionization and the two-photon ionization process used in the measurement. The phase of the transition matrix element describing a single ionization process towards a final state with angular momentum $\ell$ is the scattering phase $\eta_{\ell}$, i.e. the phase accumulated by the photoelectron when escaping from the atom. Its energy derivative, $\tau_\textsc{w} = \hbar \partial\eta_\ell\of{\epsilon}/\partial \epsilon$ represents the ``photoionization time delay" also called Wigner time delay~\cite{WignerPR1955,CarvalhoPhysRep2002}. Clearly, both $\eta_{\ell}$ and $\tau_\textsc{w}$ depend on the details of the atomic potential and their computation remains a challenge for theory.\\
Using second-order perturbation theory, the transition matrix element for two-photon ionization involving absorption of a harmonic photon $\omega_\textsc{h}$ and a laser photon $\omega$ from an initial state $\varphi_i$ to a continuum state $\varphi_{\vec{k}}$ with asymptotic momentum $\vec{k}$ can be written as
\begin{equation}
M^{(2)}_a\of{\vec{k}} =-i E_\textsc{l} E_\textsc{h} \lim_{\varepsilon\rightarrow 0^+} \sumint_n \frac{\langle\varphi_{\vec{k}}|\vec{\epsilon}\cdot\vec{r}|\varphi_{n}\rangle\langle\varphi_{n}|\vec{\epsilon}\cdot\vec{r}|\varphi_{i}\rangle}{\epsilon_i+\omega_\textsc{h}-\epsilon_n+i\varepsilon}.
\label{M2}
\end{equation}
Atomic units are used throughout. The complex amplitudes of the laser and harmonic fields are denoted $E_\textsc{l}$ and $E_\textsc{h}$ and $\vec{\epsilon}$ is their common polarization vector. The energies of the initial and intermediate states are denoted $\epsilon_i$ and $\epsilon_n$, respectively. The integral-sum is performed over all possible intermediate states $\varphi_n$. The index $a$ indicates that we first discuss a two-photon process with absorption of the laser photon.\\
We consider the channels $s \rightarrow p \rightarrow \ell$ with $\ell=s,d$. Using spherical coordinates, separating radial and angular parts of the wave functions, and expanding the final wave function into partial waves, the transition matrix element becomes
\begin{equation}
M^{(2)}_a\of{\vec{k}} =-i E_\textsc{l} E_\textsc{h} \sum_{\ell=0,2} C_{\ell0} Y_{\ell0}\of{\hat{k}} e^{i\eta_{\ell}(k)} T^{(2)}_a\of{k},
\label{M3}
\end{equation}
where $Y_{\ell 0}$ is a spherical harmonic, $C_{\ell0}$ the corresponding angular coefficient, and $\eta_{\ell}$ the scattering phase of the final state. The radial two-photon transition matrix element $T^{(2)}_a\of{k}$ can be expressed as~\cite{VeniardPRA1996,TomaJPB2002}
\begin{equation}
T^{(2)}_a\of{k} =
\sumint_n \frac{\langle R_{k\ell}|r|R_{n1}\rangle\langle R_{n1}|r|R_{i0}\rangle}{\epsilon_i+\omega_\textsc{h}-\epsilon_n+i\varepsilon}=
\langle R_{k\ell}|r|\rho_{k_a1}\rangle.
\label{introperturbedfunction}
\end{equation}
In the right member of equation~(\ref{introperturbedfunction}) we introduce the {\it perturbed} wave function $\rho_{k_a1}$ with the wave number  $k_a$ such that $k_{a}^2/2=\epsilon_i+\omega_\textsc{h}={k}^2/2-\omega$ (see Fig.~\ref{fig:1})\cite{DalgarnoPRS1955}.
To get an estimate of the phase of $T^{(2)}_a$, we consider the asymptotic behavior of the wave functions involved in equation~(\ref{introperturbedfunction}). The perturbed wave function $\rho_{k_a1}$ is an outgoing wave~\cite{AymarJPB1980,EdwardsPRA1987}
\begin{equation}
\lim_{r \rightarrow \infty}{\rho_{k_a1}(r)} \propto e^{i\left[k_ar-\tfrac{1}{2} \pi+\tfrac{1}{k_a}\ln(2k_ar)+\eta_1(k_a)\right]},
\label{a1}
\end{equation}
while $R_{k\ell}$ is real with an asymptotic behavior:
\begin{equation}
\lim_{r \rightarrow \infty} R_{k\ell}(r)  \propto \sin\left[kr-\tfrac{\ell}{2} \pi+\tfrac{1}{k}\ln(2kr)+\eta_\ell(k)\right].
\label{a2}
\end{equation}
The factor $\ell\pi/2$ arises from the centrifugal potential, while $\ln(2kr)/k$ is a correction due to the long range Coulomb potential. Using Eqs.~(\ref{M3}) - (\ref{a2}) we find an approximate expression for $M^{(2)}_a\of{k}$
\begin{equation}
M^{(2)}_a\of{k} \propto \underbrace{e^{i\eta_1(k_a)}}_{\rm(I)}\times
\underbrace{\left(\frac{i}{k_a-k}\right)^{iz}\,\,
\frac{(2k)^{\tfrac{i}{k}}}{(2k_a)^{\tfrac{i}{k_a}}}\,
\Gamma(2+iz)}_{\rm(II)},
\label{M_final}
\end{equation}
where $z=1/k_a-1/k$ and $\Gamma(z)$ is the complex Gamma function.
The first phase term (I) is the scattering phase of the intermediate state and identical to the phase of the corresponding one-photon ionization. The phase of term (II) can be assigned to the laser-driven transition connecting the two continuum states in the presence of the long-range Coulomb potential, $\varphi_a^{\textsc{cc}}$. It is independent of the short range behavior of the atomic potential and therefore {\it universal}. Corrections to this approximation due to the core are expected to become important only at energies close to threshold.\\
The phase of the two-photon matrix element $M^{(2)}_e$ for the second pathway, i.e. absorption of an harmonic photon $\omega_\textsc{h}$ followed by emission of an infrared photon $\omega$ via an intermediate state with wave number $k_e^2/2=k^2/2+\omega$  (see Fig.~\ref{fig:1}), can be derived in a similar manner. The total interference signal is obtained by angular integration of $|M^{(2)}_a + M^{(2)}_e|^2$. It can be written as Eq.~(\ref{eq:RABITT}), with
\begin{equation}
\tau_\textsc{i}= \underbrace{ \frac{\eta_1(k_e)-\eta_1(k_a)}{2\omega}}_{\tau_\textsc{w}}+\underbrace{\frac{\varphi_{e}^{\textsc{cc}}(k)-\varphi_{a}^{\textsc{cc}}(k)}{2\omega}}_{\tau_{\textsc{cc}}}.
\label{taui}
\end{equation}
This result gives an intuitive understanding of the ionization time $\tau_\textsc{i}(3s)$. It can be expressed as the sum of the Wigner time delay $\tau_\textsc{w}$ for one-photon ionization  $3s\rightarrow \epsilon p$ and an additional continuum-continuum delay $\tau_{\textsc{cc}}$ inherent to the measuring process. This analytical derivation can be easily generalized to other ionization channels.\\
Fig.~\ref{fig:3} shows the delays involved in the three ionization channels $3p\rightarrow \epsilon s$ (a), $3p\rightarrow \epsilon d$ (b) and $3s\rightarrow \epsilon p$ (c) in Ar as a function of kinetic energy.  The Wigner time delay $\tau_{\textsc{w}}$ (red) is obtained by taking the derivative of the scattering phase taken from~\cite{KennedyPRA1972}. For comparison, we also show in (d) the delays for the pathway $1s \rightarrow \epsilon p$ in hydrogen in the same energy region. The continuum-continuum delay $\tau_{\textsc{cc}}$ (blue) is calculated for a 800 nm laser wavelength and identical for all the channels and atoms. The black line indicates  $\tau_\textsc{i}$ as the sum of the two contributions.
The Wigner time delay variation can be nicely and intuitively interpreted. Low-energy electrons take longer time to escape from a given subshell than high-energy electrons. Furthermore, electrons escaping to a channel with higher angular momentum take longer time than those escaping towards a channel with low angular momentum because of the centrifugal barrier. The continuum-continuum delay has the opposite behavior and leads to an {\it apparent} quicker escape for the low-energy electrons. Finally, we also indicate in Fig.~\ref{fig:3}(d) results from exact calculations in H (dashed line). The comparison between the black and dashed lines gives an estimation of the error made in considering only the asymptotic behaviors of the perturbed and final wave functions.\\
For the energy range considered in the present work the asymmetry parameter remains close to two~\cite{HoulgateJESRP1976}, which indicates that the ionization channel $3p\rightarrow \epsilon d$ dominates over  $3p\rightarrow \epsilon s$. Neglecting the $3p\rightarrow \epsilon s$ channel, we calculate $\tau_\textsc{i}(3s) - \tau_\textsc{i}(3p)$ at the same excitation energy, using the approximation presented in Eq.~(\ref{taui}). Fig.~\ref{fig:4} presents the approximated delays (black line), together with the experimental results ($\diamond$).
We also show the Wigner time delay (red line) and the influence of the laser-driven continuum-continuum transition (blue line). The experimental results at the two highest energies agree well with the results of our calculation. The lowest energy point, however, lies outside our estimated uncertainty.
In this region the core plays a more important role for the continuum-continuum transition, as well as the Wigner time delays may differ from those calculated in \cite{KennedyPRA1972}. Using our estimated continuum-continuum delays, we can tentatively deduce the difference in photoemission delays to be equal to 140~as at 34~eV and -20~as at 37 and 40~eV.

\noindent In conclusion, we have performed experimental measurements of photoemission from the $3s^2$ and $3p^6$ shells in Ar, using a weak laser field to probe the created electron wave packets by interferometry. We identify two contributions to the measured delays: the Wigner time delay, which represents the group delay of the electron wave packet emitted in single photon ionization, and a delay inherent to the measurement process. It can be simply estimated using an universal formula which only depends on the laser frequency. We believe that the work presented here will stimulate numerous experiments, aiming at measuring photoemission delays in relative or possibly absolute values, in a variety of systems. The recent development of high-order harmonic generation using mid-infrared drivers~\cite{ChenPRL2010,AgostiniContempPhys2008} should lead both to a higher spectral resolution for this type of measurements as well as to a broader investigation bandwidth.\\
We thank Stefan Haessler, Franck L\'epine and Kenneth J. Schafer for stimulating discussions. This research was supported by the European Research Council (ALMA), the Marie Curie Intra-European Fellowship ATTOCO, the Knut and Alice Wallenberg Foundation, the Joint Research Programme ALADIN of Laserlab-Europe II and the Swedish Research Council. We acknowledge financial support from the ANR ATTO-WAVE. P.J. acknowledges support from the Swedish Foundation for Strategic Research.


\begin{thebibliography}{12}

\bibitem{SchmidtRPP1992}
V.~Schmidt,
\emph{Rep. Prog. Phys.} {\bf 55,} 1483 (1992).


\bibitem{CavalieriNature2007}
A.L.~Cavalieri {\it et~al.},
\emph{Nature} {\bf 449,} 1029 (2007).

\bibitem{SansoneScience2006}
G.~Sansone {\it et~al.},
\emph{Science} {\bf 314,} 443 (2006).

\bibitem{GoulielmakisScience2008}
E.~Goulielmakis {\it et~al.},
\emph{Science} {\bf 320,}  1614 (2008).

\bibitem{SchultzeScience2010}
M.~Schultze {\it et~al.},
\emph{Science} {\bf 328,} 1658 (2010).

\bibitem{YakovlevPRL2010}
V.S.~Yakovlev, J.~Gagnon, N.~Karpowicz, and F.~Krausz,
\emph{Phys. Rev. Lett.} {\bf 105,} 073001 (2010).

\bibitem{BaggesenPRL2010}
J.C.~Baggesen and L.B.~Madsen,
\emph{Phys. Rev. Lett.} {\bf 104,} 043602 (2010).

\bibitem{ZhangPRA2010}
C.-H.~Zhang and U.~Thumm,
\emph{Phys. Rev. A} {\bf 82,} 043405 (2010).

\bibitem{KheifetsPRL2010}
A.S.~Kheifets and I.A.~Ivanov,
\emph{Phys. Rev. Lett.} {\bf 105,} 233002 (2010).

\bibitem{WignerPR1955}
E.P.~Wigner,
\emph{Phys. Rev.} {\bf 98,} 145 (1955).

\bibitem{CarvalhoPhysRep2002}
C.A.A.~de Carvalho and H.M.~Nussenzveig,
\emph{Phys. Rep.} {\bf 364,} 83 (2002).

\bibitem{SchinsJOSA1996}
J.M.~Schins {\it et al.},
\emph{J. Opt. Soc. Am.} {\bf B13,} 197 (1996).

\bibitem{VeniardPRA1996}
V.~V\'eniard, R.~Ta\"ieb, and A.~Maquet,
\emph{Phys. Rev.~A }{\bf 54,} 721 (1996).

\bibitem{PaulScience2001}
P.M.~Paul {\it et~al.},
\emph{Science }{\bf 292,} 1689 (2001).

\bibitem{TomaJPB2002}
E.S.~Toma, and H.G.~Muller,
\emph{J. Phys.} {\bf B 35,} 3435 (2002).

\bibitem{HaesslerPRA2009}
S.~Haessler {\it et~al.},
\emph{Phys. Rev.~A} {\bf 80,} 011404 (2009).

\bibitem{SwobodaPRL2010}
M.~Swoboda {\it et al.},
\emph{Phys. Rev. Lett.} {\bf 104,} 103003 (2010).

\bibitem{FordellOptExp2009}
T.~Fordell, M.~Miranda, A.~Persson, and A~L'Huillier,
\emph{Opt. Express} {\bf 17,} 21091 (2009).

\bibitem{LopezPRL2005}
R.\,L\'opez-Martens\,{\it et~al.},\,\emph{Phys.\,Rev.\,Lett.}\,{\bf 94,}\,033001\,(2005).

\bibitem{XRayDatabase}
http://www.cxro.lbl.gov.

\bibitem{DalgarnoPRS1955}
A.~Dalgarno and J.T.~Lewis,
\emph{Proc. R. Soc., A} \textbf{233,} 70 (1955).

\bibitem{AymarJPB1980}
M.~Aymar and M.~Crance,
\emph{J. Phys. B} \textbf{30,} L 287 (1980).

\bibitem{EdwardsPRA1987}
M.~Edwards, X.~Tang, and R.~Shakeshaft,
\emph{Phys. Rev. A} {\bf 35,} 3758 (1987).

\bibitem{KennedyPRA1972}
D.J.~Kennedy and S.T.~Manson,
\emph{Phys. Rev. A} {\bf 5,} 227 (1972).

\bibitem{HoulgateJESRP1976}
R.G.~Houlgate, J.B.~West, K.~Codling, and G.V.~Marr,
\emph{J. Electr. Spect.} {\bf 9,} 205 (1976).


\bibitem{ChenPRL2010}
M.C.~Chen {\it et~al.},
\emph{Phys. Rev. Lett.} {\bf 105,} 173901 (2010).

\bibitem{AgostiniContempPhys2008}
P.~Agostini and L.F.~DiMauro,
\emph{Contemp. Phys.} {\bf 49,} 179 (2008).

\end{thebibliography}
\end{document}